\documentclass[aps,prl,preprint,nopacs,superscriptaddress]{revtex4}
\usepackage{amsmath}
\usepackage{amssymb}
\usepackage{graphicx}
\usepackage{hyperref}
\newcommand{\beq}{\begin{equation*}}
\newcommand{\eeq}{\end{equation*}}

\begin{document}

\title{Observation of surface states derived from topological Fermi arcs in the Weyl semimetal NbP}

\author{Ilya Belopolski$^*$} \affiliation{Laboratory for Topological Quantum Matter and Spectroscopy (B7), Department of Physics, Princeton University, Princeton, New Jersey 08544, USA}
\author{Su-Yang Xu$^*$} \affiliation{Laboratory for Topological Quantum Matter and Spectroscopy (B7), Department of Physics, Princeton University, Princeton, New Jersey 08544, USA}
\author{Daniel Sanchez\footnote{These authors contributed equally to this work.}} \affiliation{Laboratory for Topological Quantum Matter and Spectroscopy (B7), Department of Physics, Princeton University, Princeton, New Jersey 08544, USA}
\author{Guoqing Chang} \affiliation{Centre for Advanced 2D Materials and Graphene Research Centre, National University of Singapore, 6 Science Drive 2, Singapore 117546} \affiliation{Department of Physics, National University of Singapore, 2 Science Drive 3, Singapore 117542}
\author{Cheng Guo} \affiliation{International Center for Quantum Materials, Peking University, Beijing 100871, China}
\author{Madhab Neupane} \affiliation{Condensed Matter and Magnet Science Group, Los Alamos National Laboratory, Los Alamos, NM 87545, USA}
\author{Hao Zheng} \affiliation{Laboratory for Topological Quantum Matter and Spectroscopy (B7), Department of Physics, Princeton University, Princeton, New Jersey 08544, USA}
\author{Chi-Cheng Lee} \affiliation{Centre for Advanced 2D Materials and Graphene Research Centre, National University of Singapore, 6 Science Drive 2, Singapore 117546} \affiliation{Department of Physics, National University of Singapore, 2 Science Drive 3, Singapore 117542}
\author{Shin-Ming Huang} \affiliation{Centre for Advanced 2D Materials and Graphene Research Centre, National University of Singapore, 6 Science Drive 2, Singapore 117546} \affiliation{Department of Physics, National University of Singapore, 2 Science Drive 3, Singapore 117542}
\author{Guang Bian} \affiliation{Laboratory for Topological Quantum Matter and Spectroscopy (B7), Department of Physics, Princeton University, Princeton, New Jersey 08544, USA}
\author{Nasser Alidoust} \affiliation{Laboratory for Topological Quantum Matter and Spectroscopy (B7), Department of Physics, Princeton University, Princeton, New Jersey 08544, USA}
\author{Tay-Rong Chang} \affiliation{Laboratory for Topological Quantum Matter and Spectroscopy (B7), Department of Physics, Princeton University, Princeton, New Jersey 08544, USA} \affiliation{Department of Physics, National Tsing Hua University, Hsinchu 30013, Taiwan}
\author{BaoKai Wang} \affiliation{Centre for Advanced 2D Materials and Graphene Research Centre, National University of Singapore, 6 Science Drive 2, Singapore 117546} \affiliation{Department of Physics, National University of Singapore, 2 Science Drive 3, Singapore 117542} \affiliation{Department of Physics, Northeastern University, Boston, Massachusetts 02115, USA}
\author{Xiao Zhang} \affiliation{International Center for Quantum Materials, Peking University, Beijing 100871, China}
\author{Arun Bansil} \affiliation{Department of Physics, Northeastern University, Boston, Massachusetts 02115, USA}
\author{Horng-Tay Jeng} \affiliation{Department of Physics, National Tsing Hua University, Hsinchu 30013, Taiwan} \affiliation{Institute of Physics, Academia Sinica, Taipei 11529, Taiwan}
\author{Hsin Lin} \affiliation{Centre for Advanced 2D Materials and Graphene Research Centre, National University of Singapore, 6 Science Drive 2, Singapore 117546} \affiliation{Department of Physics, National University of Singapore, 2 Science Drive 3, Singapore 117542}
\author{Shuang Jia} \affiliation{International Center for Quantum Materials, Peking University, Beijing 100871, China}
\author{M. Zahid Hasan} \affiliation{Laboratory for Topological Quantum Matter and Spectroscopy (B7), Department of Physics, Princeton University, Princeton, New Jersey 08544, USA} \affiliation{Princeton Center for Complex Materials, Princeton Institute for Science and Technology of Materials, Princeton University, Princeton, New Jersey 08544, USA}


\begin{abstract}
The recent experimental discovery of a Weyl semimetal in TaAs provides the first observation of a Weyl fermion in nature and demonstrates a novel type of anomalous surface state band structure, consisting of Fermi arcs. So far, work has focused on Weyl semimetals with strong spin-orbit coupling (SOC). However, Weyl semimetals with weak SOC may allow tunable spin-splitting for device applications and may exhibit a crossover to a spinless topological phase, such as a Dirac line semimetal in the case of spinless TaAs. NbP, isostructural to TaAs, may realize the first Weyl semimetal in the limit of weak SOC. Here we study the surface states of NbP by angle-resolved photoemission spectroscopy (ARPES) and we find that we $\textit{cannot}$ show Fermi arcs based on our experimental data alone. We present an $\textit{ab initio}$ calculation of the surface states of NbP and we find that the Weyl points are too close and the Fermi level is too low to show Fermi arcs either by (1) directly measuring an arc or (2) counting chiralities of edge modes on a closed path. Nonetheless, the excellent agreement between our experimental data and numerical calculations suggests that NbP is a Weyl semimetal, consistent with TaAs, and that we observe trivial surface states which evolve continuously from the topological Fermi arcs above the Fermi level. Based on these results, we propose a slightly different criterion for a Fermi arc which, unlike (1) and (2) above, does not require us to resolve Weyl points or the spin splitting of surface states. We propose that raising the Fermi level by $> 20$ meV would make it possible to observe a Fermi arc using this criterion in NbP. Our work offers insight into Weyl semimetals with weak spin-orbit coupling, as well as the crossover from the spinful topological Weyl semimetal to the spinless topological Dirac line semimetal.
\end{abstract}

\date{\today}
\maketitle

A Weyl semimetal is a crystal which hosts Weyl fermions as emergent quasiparticles \cite{Weyl, Herring, Abrikosov, Nielsen, Volovik, Murakami, Pyrochlore, Multilayer}. Although Weyl fermions are well-studied in quantum field theory, they have not been observed as a fundamental particle in nature. The recent experimental observation of Weyl fermions in TaAs offers a beautiful example of emergence in science and grants access in experiment to a wealth of phenomena associated with Weyl fermions in theory \cite{TaAsThyUs, TaAsThyThem, TaAsUs, LingLu, TaAsThem}. Weyl semimetals also give rise to a topological classification which is closely related to the Chern number of the integer quantum Hall effect \cite{Vish}. In the bulk band structure of a three-dimensional sample, Weyl fermions correspond to points of accidental degeneracy, Weyl points, between two bands. The Chern number on a two-dimensional slice of the Brillouin zone passing in between Weyl points can be non-zero. In addition, this Chern number changes when the slice is swept through a Weyl point. As in the quantum Hall effect, the Chern number in a Weyl semimetal protects topological boundary modes. However, because the Chern number changes when the slice is swept through a Weyl point, the boundary modes have the exotic property that they terminate in momentum space at the locations of Weyl points. On the two-dimensional surface of a Weyl semimetal, the resulting surface state band structure consists of topological Fermi arcs, with constant energy contours which do not form closed curves. In this way, Weyl semimetals provide the most dramatic example to date of an anomalous surface state band structure. While the first experiments on Weyl semimetals studied compounds with strong spin-orbit coupling (SOC), it is unclear how Fermi arcs in a Weyl semimetal evolve in the limit of weak SOC \cite{TaAsChen, TaAsNodesDing, NbAs, TaPUs, TaPThem}. Moreover, if SOC is ignored, \textit{ab initio} numerical calculations predict that TaAs realizes a novel phase known as a spinless topological Dirac line semimetal, with a touching of four bands along an entire closed curve in the bulk Brillouin zone \cite{BurkovLine}. NbP is isostructural to TaAs and is predicted to be a Weyl semimetal \cite{TaAsThyUs, FourCompounds}. However, NbP has much lower atomic number, placing it in the limit of weak SOC. As a result, NbP provides the first known example in theory of a Dirac line semimetal and further offers the opportunity to understand the crossover from a spinful Weyl semimetal to a spinless Dirac line semimetal. The crossover to a spinless system is also associated with a tunable spin-splitting in the bulk and surface band structure, which may be useful in applications. The low spin-orbit coupling in NbP, the topologically non-trivial spinless phase of NbP and the availability of TaAs as a spin-orbit coupled cousin to NbP offer the chance to realize a Dirac line semimetal, study a crossover to a Weyl semimetal and understand the evolution of Fermi arcs in the limit of weak spin-orbit coupling.

Here we use vacuum ultraviolet angle-resolved photoemission spectroscopy (ARPES) to study the surface state band structure of the (001) surface of NbP. We do not directly observe topological Fermi arc surface states in the sense that all surface state constant-energy contours form closed curves. We are also unable to establish Fermi arcs by counting chiralities of edge modes on a loop in momentum space, a trick which has recently been applied to demonstrate a non-zero Chern number in TaP \cite{TaPUs}. However, we present an \textit{ab initio} calculation of the (001) surface states of NbP and find that we can reproduce with excellent agreement all trivial surface states. Our calculation further shows that the spin-splitting in NbP is very small due to the low spin-orbit coupling, reducing the separation between Weyl points. In addition, we find that the most well-separated Weyl points in NbP are well above the Fermi level. Both of these facts make it difficult to demonstrate Fermi arcs by ARPES on the (001) surface of NbP. Nonetheless, the excellent agreement between our \textit{ab initio} calculations and our ARPES spectra suggests that NbP is a Weyl semimetal, consistent with TaAs, NbAs and TaP. Our calculations further relate a trivial surface state at the Fermi level to a Fermi arc $\sim 26$ meV above the Fermi level. In this sense, our ARPES spectra show trivial surface states which evolve continuously into topological Fermi arcs above the Fermi level. We point out that if these states were accessible, then an ARPES experiment could demonstrate that such a contour is a Fermi arc without resolving either the two Weyl points to which it is attached or spin-splitting in the Fermi arc. Instead, it would be sufficient to consider the evolution of the surface states in binding energy. We emphasize that such an approach to demonstrating Fermi arcs is distinct from both a direct observation of an open constant-energy contour and the counting of chiralities of edge modes. We summarize these three distinct ways of showing Fermi arcs in a Weyl semimetal. Lastly, we discuss spin-splitting in our ARPES data and we show that the trivial surface states have contributions from multiple atomic orbitals, giving rise to different surface state pockets which we observe do not hybridize. Our work provides insight into an inversion-breaking Weyl semimetal in the limit of a Dirac line semimetal, without spin-orbit coupling. Our work also offers a useful summary of the ways in which we can demonstrate in general Fermi arcs in Weyl semimetals.

We first present the compound under study. Niobium phosphide (NbP) crystallizes in a body-centered tetragonal Bravais lattice, in point group $C_{4v}$ ($4mm$), space group $I4_1md$ (109), isostructural to TaAs, TaP and NbAs \cite{Crystal1, Crystal2, Crystal3}. The crystal structure can be understood as a stack of alternating Nb and P square lattice layers, see Fig. \ref{Fig1}a. Each layer is shifted with respect to the one below it by half an in-plane lattice constant, $a/2$, in either the $\hat{x}$ or $\hat{y}$ direction. The crystal structure can also be understood as arising from intertwined helices of Nb and P atoms which are copied in-plane to form square lattices, with one Nb (or P) atom at every $\pi/2$ rad along the helix. The conventional unit cell consisting of one period of the helices is shown in Fig. \ref{Fig1}b. This helical structure is related to the non-symmorphic $C_4$ symmetry, where a $C_4$ rotation followed by a translation by $c/4$ is required to take the crystal back into itself. We note that NbP has no inversion symmetry, so that all bands are generically singly-degenerate. This is a crucial requirement for NbP to be a Weyl semimetal. We show a photograph of the sample taken through an optical microscope, suggesting that it is a single crystal and of high quality, in Fig. \ref{Fig1}c. A scanning tunneling microscopy (STM) topography of the sample shows a square lattice surface, demonstrating that NbP cleaves on the (001) plane, see Fig. \ref{Fig1}d. The lack of defects further suggests the high quality of the single crystals. From the ionic model, we expect that the conduction and valence bands in NbP arise from Nb $4d$ and P $3p$ orbitals, respectively. However, an \textit{ab initio} bulk band structure calculation along high-symmetry lines shows that NbP does not have a full gap but is instead a semimetal, see $\Sigma-\Gamma$, $Z-\Sigma'$, $\Sigma'-N$ in Fig. \ref{Fig1}e, with the bulk Brillouin zone in Fig. \ref{Fig1}f. In the absence of spin-orbit coupling, the band structure near the Fermi level consists of four Dirac lines, shown in purple in Fig. \ref{Fig1}g. These Dirac lines are protected by two vertical mirror planes, shown in blue. After spin-orbit coupling is included, each Dirac line vaporizes into six Weyl points shifted slightly off the mirror plane, marked by the black and white dots in Fig. \ref{Fig1}g. Two Weyl points are on the $k_z = 0$ plane, shown in red, and we call these Weyl points $W_1$. The other four, we call $W_2$. We note that on the (001) surface, two $W_2$ of the same chirality project onto the same point of the surface Brillouin zone, giving rise to a projected Weyl point of chiral charge $\pm 2$. The $W_1$ give projections of chiral charge $\pm 1$, see Fig. \ref{Fig1}h.

On the basis of our \textit{ab initio} results, we search for Weyl points and Fermi arcs in NbP by ARPES. First, we show that we observe surface states but not bulk states in vacuum ultraviolet ARPES on the (001) surface of NbP. In our ARPES spectra, we observe a Fermi surface consisting of lollipop-shaped pockets along the $\bar{\Gamma} - \bar{X}$ and $\bar{\Gamma} - \bar{Y}$ lines and peanut-shaped pockets on the $\bar{M} - \bar{X}$ and $\bar{M} - \bar{Y}$ lines, see Fig. \ref{Fig2}a-e. The spectra are consistent with the other compounds in the same family, suggesting that these pockets are surface states rather than bulk states. Because $C_4$ symmetry is implemented as a screw axis in NbP, the (001) surface breaks $C_4$ symmetry and the surface state dispersion is not $C_4$ symmetric. Our data suggest that the peanut pockets at $\bar{X}$ and $\bar{Y}$ differ slightly, showing that we observe surface states. We note, however, that this effect is much weaker than in TaAs or TaP \cite{TaAsUs, TaPUs}. This result could be explained by reduced coupling between the square lattice layers, restoring the $C_4$ symmetry of each individual layer. In particular, we note that the lattice constants of NbP are comparable to those of TaAs, while the atomic orbitals are smaller due to the lower atomic number. We expect this effect to be particularly important for surface states derived from the $p_x$, $p_y$, $d_{xy}$ and $d_{x^2-y^2}$ orbitals and indeed we observe no $C_4$ breaking at all for the lollipop pockets, which arise from the in-plane orbitals \cite{FourCompounds}. We conclude that we observe the surface states of NbP in our ARPES spectra.

Because our ARPES spectra show the (001) surface states of NbP, we ask whether we observe Fermi arcs. We see in Figs. \ref{Fig2}a-e that both the lollipop and peanut pockets are closed, so we observe no single disconnected arc. We also see no evidence of a kink in the constant-energy contours, so we do not observe a pair of arcs connecting to the same $W_2$ in a discontinuous way. It is also possible that the two arcs approach the $W_2$ with approximately the same slope. To exclude this possibility, we present a difference map of two ARPES spectra, at $E_B = 0.05$ eV and $E_B = 0.1$ eV, corresponding to the Fermi velocities at every point on each pocket, see Fig. \ref{Fig2}g. We see that the Fermi velocities have the same sign all the way around both the lollipop and the peanut pockets. If the lollipop consisted of Fermi arcs, one arc should evolve in a hole-like way, while the other arc should evolve in an electron-like way, so the different regions of the lollipop pocket should have opposite Fermi velocities in this sense. Because all points on both pockets have the same Fermi velocity, these pockets are trivial, hole-like surface states, not topological Fermi arcs. We can also consider a closed path through the surface Brillouin zone and count chiralities of edge modes along the path. We check (1) a triangular path, $\mathcal{C}$, along $\bar{\Gamma}-\bar{X}-\bar{M}-\bar{\Gamma}$, which encloses net chiral charge $+1$, Fig. \ref{Fig2}i, and (2) a small circular path, $\mathcal{P}$, which encloses net chiral charge $-2$, Fig. \ref{Fig2}j. For each path, we label each spinless crossing with an up or down arrow to indicate the sign of the Fermi velocity. We find that going around either $\mathcal{C}$ and $\mathcal{P}$ we have net zero chirality, showing zero Chern number on the associated bulk manifold. This result is reasonable because our spectra simply consist of two overlapping hole pockets, illustrated in Fig. \ref{Fig2}h.

Next, we compare our experimental results to numerical calculations on NbP and show that it is challenging to observe Fermi arcs in our spectra because of the low spin-orbit coupling. We present a calculation of the (001) surface states in NbP for the P termination, at the binding energy of $W_2$, $\varepsilon_{W2} = -0.026$ eV, at the Fermi level and at the binding energy of $W_1$, $\varepsilon_{W1} = 0.053$ eV, see Figs. \ref{Fig3}a-c. We also plot the Weyl point projections, obtained from a bulk band structure calculation \cite{FourCompounds}. We observe surface states (1) near the mid-point of the $\bar{\Gamma}-\bar{X}$ and $\bar{\Gamma}-\bar{Y}$ lines and (2) near $\bar{Y}$ and $\bar{X}$. The surface states (1) form two Fermi arcs and two closed contours at $\varepsilon_{W2}$, see Fig. \ref{Fig3}d,e. These states undergo a Lifshitz transition near $\varepsilon_F$ with the surface states (2), forming a large hole-like pocket below the Fermi level. The surface states (2) also form a large hole-like pocket. They contain within them, near the $\bar{X}$ and $\bar{Y}$ points, a short Fermi arc connecting each pair of $W_1$, see Fig. \ref{Fig3}f,g. We note the excellent agreement with our ARPES spectra, where we also see lollipops and peanuts which evolve into trivial, closed, hole-like pockets below $\varepsilon_F$. At the same time, we find in our calculation that the separation of Weyl points and the spin-splitting in the surface states is small. This result is consistent with our ARPES spectra, which do not show spin-splitting in the surface states near the Fermi level. We further find that the connectivity of Fermi arcs in the calculation is consistent with the topological classification even as the spin degeneracy is restored. In particular, two Fermi arcs connect to the $W_2$, but each Fermi arc is also paired with a nearly-degenerate closed contour, reflecting the low spin-splitting, see Figs. \ref{Fig3}d,e. Similarly, each pair $W_1$ are connected by one Fermi arc, whose spin-reversed partner is the Fermi arc on the other side of the Brillouin zone connecting two other $W_1$, see Figs. \ref{Fig3}f,g. In this way, the surface states obey both the topological protection of the non-zero Chern number as well as the spin degeneracy approximately enforced by low spin-orbit coupling.

The small spin-splitting observed in our numerical calculations underlines the difficulty in observing topological Fermi arc surface states in NbP. The separation of the Weyl points is $< 0.01 \AA^{-1}$ for the $W_1$ and $< 0.02 \AA^{-1}$ for the $W_2$, both well below the typical linewidth of our ARPES spectra, $\sim 0.05\AA^{-1}$. For this reason, we cannot resolve the momentum space region between the $W_1$ or the $W_2$ to determine if there is an arc. We emphasize that we cannot surmount this difficulty by considering Fermi level crossings on $\mathcal{P}$ or $\mathcal{C}$, as shown in Fig. \ref{Fig1}f. It is obvious that if we cannot resolve the two Weyl points in a Fermi surface mapping, then we also cannot resolve a Fermi arc connecting them in a cut passing through the Weyl points. In this way, on $\mathcal{P}$ we cannot verify the arc connecting the $W_1$ and on $\mathcal{P}$ and $\mathcal{C}$ we cannot verify the empty region between the $W_2$. We point out that $\mathcal{P}$ fails for TaAs, TaP, NbAs and NbP due to the small separation of the $W_1$, despite recent claims that this path can be used to demonstrate a Weyl semimetal in TaAs and NbP \cite{TaAsThem, TaAsChen, NbPThem}.

As an additional complication, it is difficult to use $\mathcal{P}$ or $\mathcal{C}$ because the Fermi level is below the Lifshitz transition for the $W_2$ in NbP, see Fig. \ref{Fig4}a. This invalidates any counting of chiralities of edge modes because for $\varepsilon_F < \varepsilon_L$, there is no accessible binding energy where the bulk band structure is gapped along an entire loop passing in between a pair of $W_2$, illustrated in Fig. \ref{Fig4}b by the broken dotted red line for $\varepsilon = \varepsilon_F$. On the bright side, if we could access $\varepsilon > \varepsilon_L$, then it may be possible to demonstrate a Fermi arc in NbP without resolving the $W_2$ and counting chiralities. In particular, while the Fermi arc may appear to form a closed contour due to the small separation of $W_2$, it could have a kink at the location of the $W_2$. The Fermi arc would also tend to disperse in one direction with binding energy, in sharp contrast to a closed contour, which would tend to grow or shrink in all directions. Unlike the previous criteria which we considered to show Fermi arcs, this criterion does not depend on resolving the $W_2$ or the spin-splitting of the surface states. Lastly, we note that we can understand the lollipop pocket at the Fermi level as arising from the outer Fermi arc and trivial surface state at $\varepsilon_{W2}$, see again Fig. \ref{Fig3}a. If we start from $\varepsilon_{W2}$ and scan to deeper binding energies, we see that both the arc and trivial surface state enlarge and eventually undergo a surface state Lifshitz transition with the $\bar{X}$ and $\bar{Y}$ pockets at $\varepsilon \sim \varepsilon_F$, see again Fig. \ref{Fig3}b. Below this Lifshitz transition, the lollipop pocket becomes a trivial hole-like pocket with approximate spin degeneracy, as shown in Fig. \ref{Fig3}c and illustrated in Fig. \ref{Fig2}h. In this sense, although we do not observe a non-zero Chern number by counting chiralities of edges modes in NbP, the excellent correspondence between our ARPES spectra and \textit{ab initio} calculation suggests that the trivial lollipop surface state evolves continuously into the topological Fermi arcs above the Fermi level.

We point out several other features of the (001) surface states of NbP. First, we note that we can observe a spin splitting in the lollipop pocket at $E_B \sim 0.2$ eV, shown in Fig. \ref{Fig4}c and repeated with guides to the eye in Fig. \ref{Fig4}d. Next, we observe in our ARPES spectra that the lollipop and peanut pockets do not hybridize anywhere in the surface Brillouin zone. We present a set of dispersions near the intersection of the lollipop and peanut pockets and we find no avoided crossings, see Fig. \ref{Fig4}e for the locations of the cuts, shown in Fig. \ref{Fig4}f. As mentioned above, we attribute this effect to the different orbital character of the two pockets. In particular, the lollipop pocket arises mostly from in-plane $p$ and $d$ orbitals and the peanut pocket arises mostly from out-of-plane $p$ and $d$ orbitals. The suppressed hybridization may be related to the $C_2$ symmetry of the (001) surface. In particular, we note that the in-plane and out-of-plane orbitals transform under different representations of $C_2$. The contributions from different, unhybridized orbitals to the surface states in NbP may give rise to novel phenomena. For example, we propose that quasiparticle interference between the lollipop and peanut pockets will be supressed in an STM experiment on NbP. Also, the rich surface state structure may explain why the bulk band structure of NbP is invisible to vacuum ultraviolet ARPES. Specifically, because the surface states take full advantages of all available orbitals, there are no orbitals left near the surface to participate in the bulk band structure. Other phenomena may also arise from the rich surface state structure in NbP.

Lastly, we compare the $W_2$ and their Fermi arcs in NbP and TaAs. Unlike NbP, TaAs hosts two Fermi arcs without any additional closed surface state contours. In addition, the spin splitting of the two Fermi arcs is much larger, see Fig. \ref{Fig4}g. Similarly, the separation of Weyl points is $\sim 4$ times larger in TaAs than NbP, see Fig. \ref{Fig4}h,i. We see again that the reduced spin-orbit coupling in NbP prevents us from directly counting chiral edge modes to experimentally measure Chern numbers. In addition, as has been noted before \cite{TaPUs}, there is an apparent switching of chirality in the Weyl points of NbP and TaAs. This arises because the $W_1$ cross over into the adjacent Brillouin zone due to the fact that the Dirac line in TaAs is smaller than in NbP, as is seen in Fig. \ref{Fig4}e,f. We see that the weak SOC of NbP prevents us from resolve the separation between Weyl points, so that we cannot directly observe Fermi arcs or demonstrate a non-zero Chern number by counting chiralities of edge modes. However, we propose that even with weak SOC, if the Fermi level of NbP could be raised to the $W_2$, then it is possible to demonstrate a Fermi arc without resolving the $W_2$ or the spin splitting of the surface states. In particular, the Fermi arcs could show a kink at the $W_2$ and would tend to disperse in one direction with binding energy. This criterion could be used to demonstrate Fermi arcs in Weyl semimetals which have yet to be discovered.

\bigskip
{\bf Acknowledgements}
\bigskip

We thank Makoto Hashimoto and Donghui Lu for technical assistance with ARPES measurements at SSRL Beamline 5-4, SLAC, Menlo Park, CA, USA. We also thank Nicholas Plumb and Ming Shi for technical assistance with ARPES measurements at the HRPES endstation of the SIS beamline, Swiss Light Source, Villigen, Switzerland.

\clearpage
\begin{figure*}
\centering
\includegraphics[width=17cm, trim={70 210 70 90}, clip]{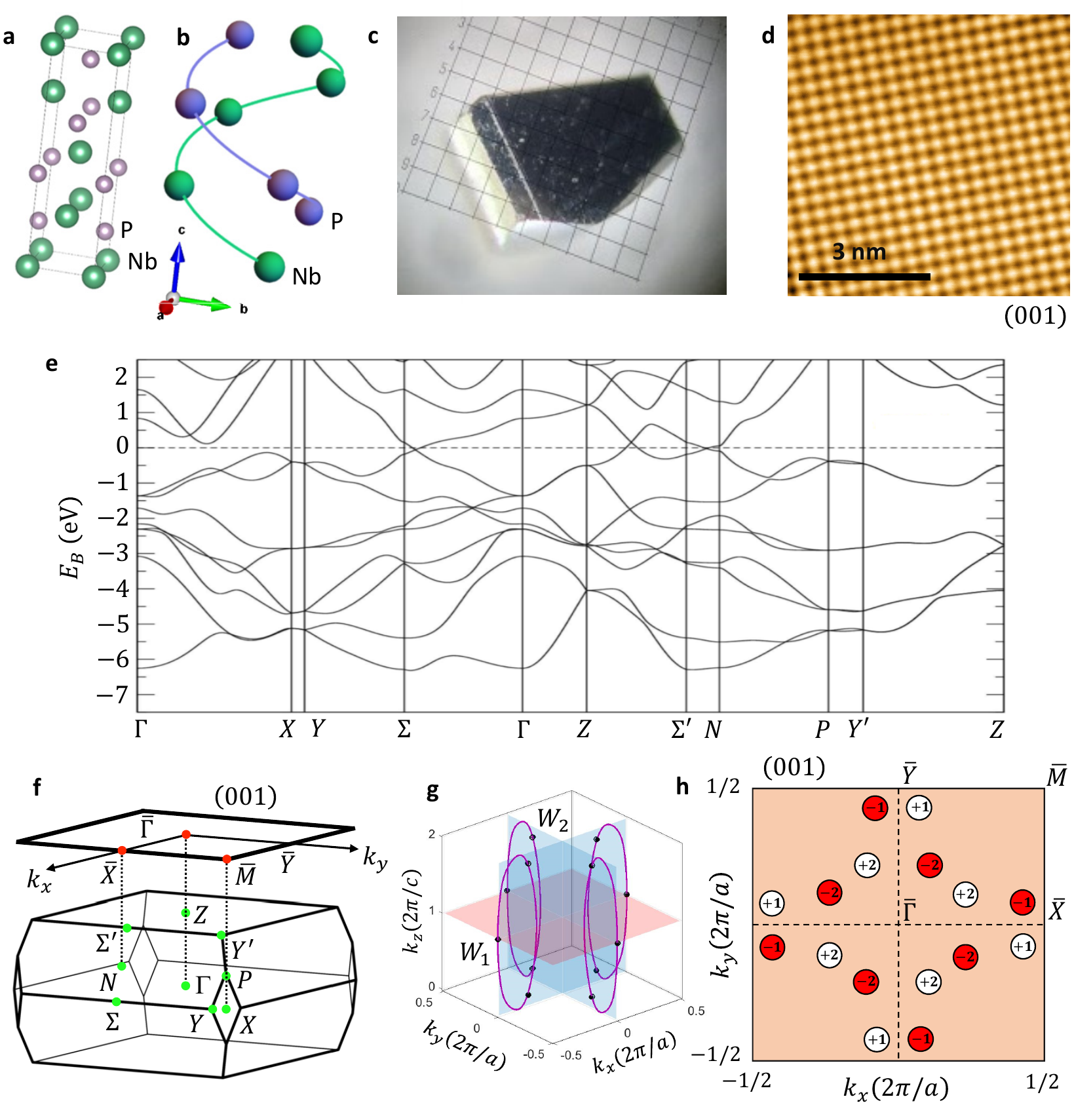}
\end{figure*}

\clearpage
\begin{figure*}
\caption{\label{Fig1}\textbf{Overview of the Weyl semimetal candidate NbP.} (a) The crystal structure of NbP, which can be understood as a stack of square lattices of Nb and P, with a stacking pattern which involves an in-plane shift of each layer relative to the one below it. (b) The crystal structure can also be understood as a pair of intertwined helices of Nb and P atoms which are copied in-plane to form square lattice layers. The axis of the helix is the $\hat{z}$ direction of the conventional unit cell, the center of the helix is $1/4$ of the way along the diagonal of one plaquette of a square lattice layer, and the radius of the helix is $a/2\sqrt{2}$. (c) Photograph of the sample taken through an optical microscope, showing a beautiful crystal. (d) An STM topography of the (001) surface of NbP, showing the high quality of the sample surface, with no defects within a 6.2 nm $\times$ 6.2 nm window. The image was taken at bias voltage $-0.3$ eV and temperature $4.6$ K. (e) \textit{Ab initio} bulk band structure calculation of NbP, using GGA exchange correlation functionals, without spin-orbit coupling (SOC), showing that NbP is a semimetal with band inversions along $\Sigma-\Gamma$, $Z-\Sigma'$ and $\Sigma'-N$. (f) The bulk Brillouin zone and (001) surface Brillouin zone of NbP, with high-symmetry points labeled. (g) Without SOC, NbP has four Dirac lines protected by two mirror planes (blue). With SOC, each Dirac line vaporizes into six Weyl points (black and white dots), two on the $k_z = 0$ plane (red), called $W_1$, and four away from $k_z = 0$, called $W_2$. (h) Illustration of the Weyl point projections in the (001) surface Brillouin zone. Two $W_2$ of the same chiral charge project onto the same point in the surface Brillouin zone, giving Weyl point projections of chiral charge $\pm 2$. The separation is not to scale, but the splitting between pairs of $W_2$ is in fact larger than the splitting between pairs of $W_1$ (see also Fig. \ref{Fig4}h).}
\end{figure*}

\clearpage
\begin{figure}
\centering
\includegraphics[width=17cm, trim={80 220 120 60}, clip]{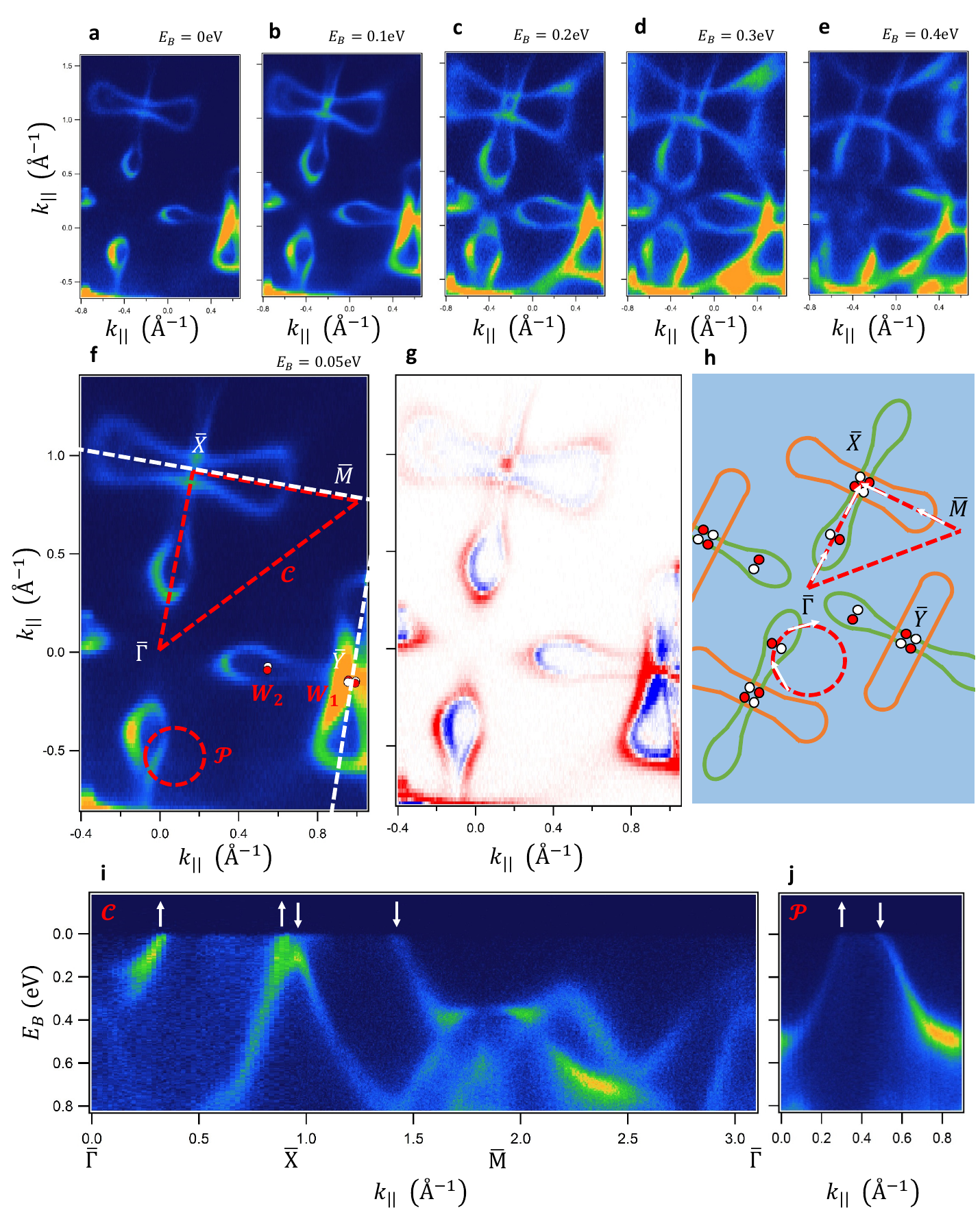}
\end{figure}

\clearpage
\begin{figure*}
\caption{\label{Fig2}\textbf{Surface states of NbP by ARPES.} (a)-(e) Fermi surface by vacuum ultraviolet APRES on the (001) surface of NbP at binding energies $E_B = 0$, $0.1$, $0.2$, $0.3$ and $0.4$ eV. The similarity with TaAs, NbAs and TaP as well as the slight $C_4$ symmetry breaking show that at vacuum ultraviolet energies the spectral weight is dominated by the surfaces states, not the bulk states of NbP. We observe lollipop and peanut-shaped pockets and find that both are hole-like. (f) Same as (a)-(e), but at $E_B = 0.05$ eV and with additional decoration to illustrate the high symmetry points of the surface Brillouin zone, the locations of the Weyl points and the two paths $\mathcal{C}$ and $\mathcal{P}$ on which we measure Chern numbers. (g) The difference of ARPES spectra at $E_B = 0.05$ eV and $E_B = 0.1$ eV, illustrating the direction of the Fermi velocity all around the lollipop and peanut pockets. The blue contour is always inside the red contour, indicating that the sign of the Fermi velocity is the same going around each contour. This result excludes the possibility that the lollipop actually consists of Fermi arcs attached to the $W_2$ in such a way as to create an apparently closed pocket. (h) Cartoon of the band structure, including the two paths $\mathcal{C}$ and $\mathcal{P}$ and the Weyl points which they enclose, and arrows indicating the chirality of each edge mode. (i) Band structure by ARPES along the path $\mathcal{C}$, with chiralities of edge modes marked by the arrows. Note that each arrow secretly corresponds to two crossings, because we cannot observe spin splitting at the Fermi level due to the weak SOC of NbP. There are the same number of arrows going up as down, so the Chern number is zero. This is easy to see because the path enters and exits two pockets. (j) Same as (i) but along the path $\mathcal{C}$. Again the Chern number is zero.}
\end{figure*}

\clearpage
\begin{figure}
\centering
\includegraphics[width=17cm, trim={120 150 60 70}, clip]{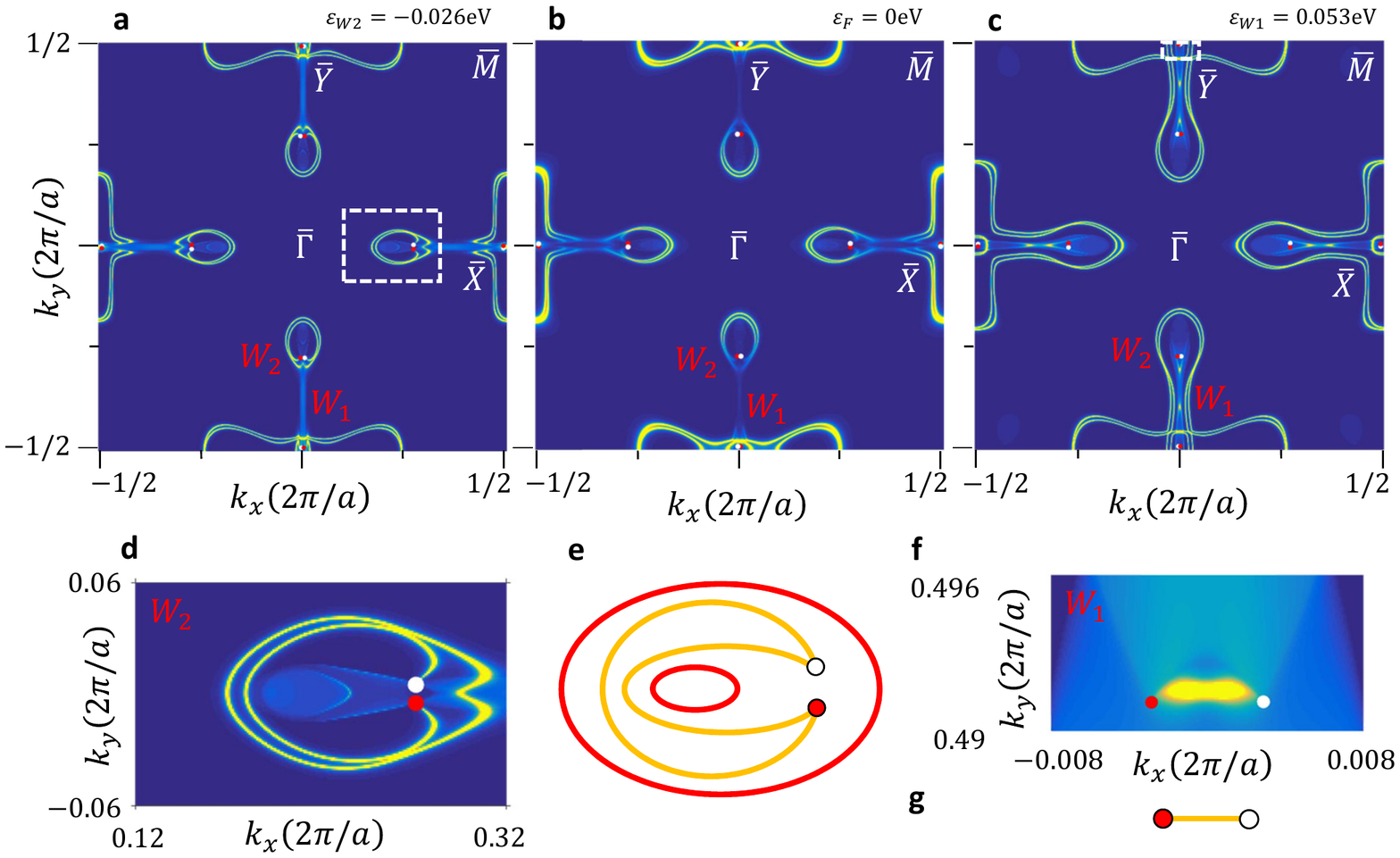}
\end{figure}

\clearpage
\begin{figure*}
\caption{\label{Fig3}\textbf{Numerical calculation of Fermi arcs in NbP.} First-principles band structure calculation of the (001) surface states of NbP at (a) the energy of the $W_2$, above the Fermi level, (b) the Fermi level and (c) the energy of the $W_1$, below the Fermi level. The projected Weyl nodes on the (001) surface are denoted by the small red and white circles. In (a) we see (1) a small set of surface states near the $W_2$ and (2) larger surface states near $\bar{X}$ and $\bar{Y}$. Near the energy of (b) there is a Lifshitz transition between the surface states at (1) and (2), giving rise to lollipop and peanut-shaped pockets. At (c) we see that the lollipop and peanut pockets enlarge, so they are hole-like. We also observe short Fermi arcs connecting the $W_1$. The numerical calculation shows excellent overall agreement with our ARPES spectra. Specifically, we find trivial, closed, hole-like lollipop and peanut pockets below the Fermi level. (d) Zoom-in of the surface states around the $W_2$, indicated by the white box in (a). We find two Fermi arcs and two trivial closed contours, illustrated in (e). (f) Zoom-in of the surface states around $W_1$, indicated by the white box at the bottom of (c). We find one Fermi arc, illustrated in (g).}
\end{figure*}

\clearpage
\begin{figure}
\centering
\includegraphics[width=17cm, trim={60 430 110 10}, clip]{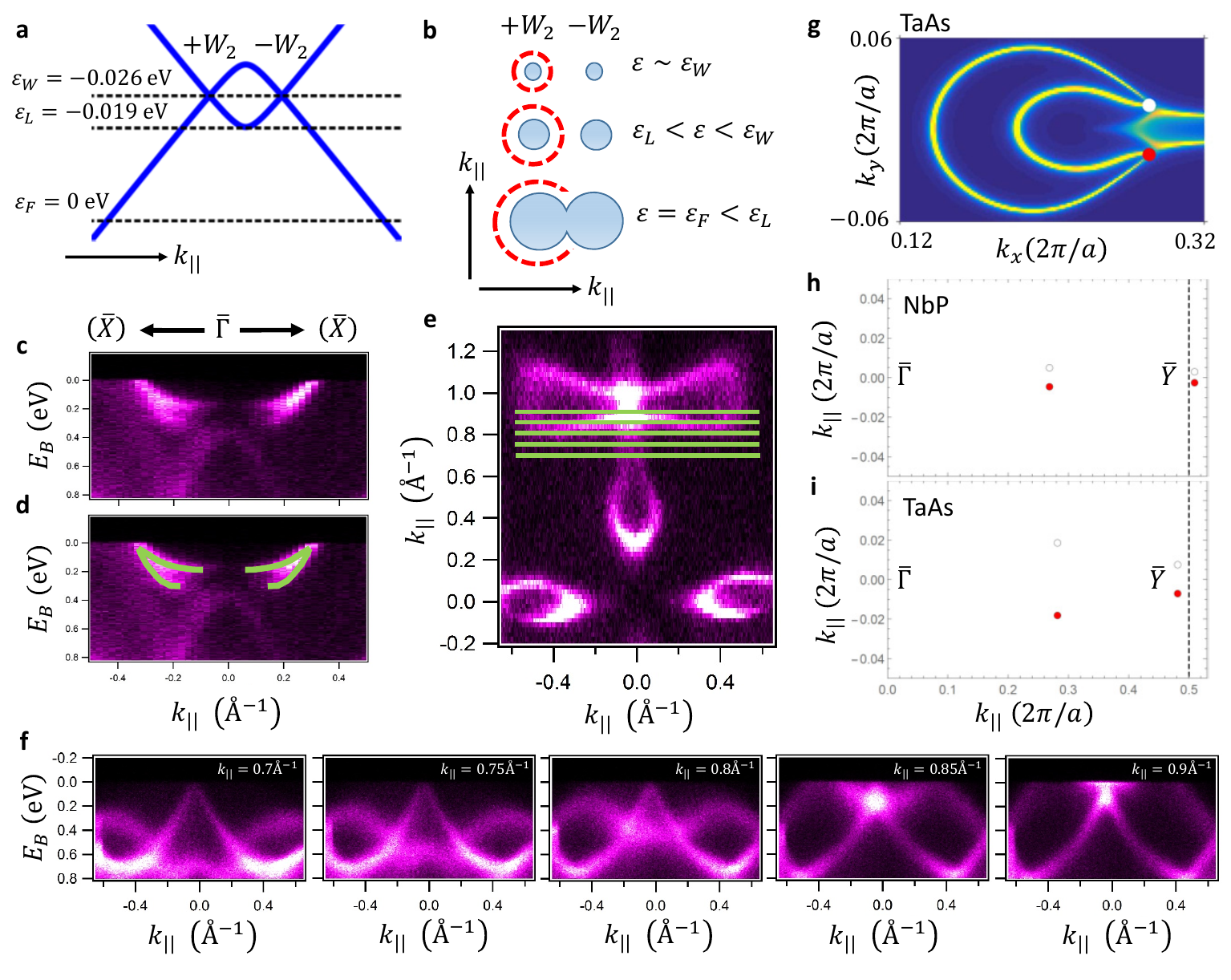}
\end{figure}

\clearpage
\begin{figure*}
\caption{\label{Fig4}\textbf{Comparison between NbP and TaAs.} (a) Relevant energies of the $W_2$ compared to the Fermi level in the numerical calculation. One consequence of the small spin-splitting is that the Lifshitz transition between the $W_2$ is only $\sim 0.007$ eV below the energy of the $W_2$ and $\sim 0.019$ eV above the Fermi level. (b) Because $\varepsilon_F < \varepsilon_L$, it makes no sense to calculate the Chern number on $\mathcal{C}$ and $\mathcal{P}$, because the two-dimensional band structure corresponding to that cut is not an insulator. This is illustrated by the interrupted dotted red line in the last row of (b). (c) Surface states by ARPES along $\bar{X}-\bar{\Gamma}-\bar{X}$, showing a spin splitting below the Fermi level. (d) Same as (c) but with guides to the eye to mark the spin splitting. (e). Fermi surface by APRES at $E_B = 0.1$ eV, marking the cuts shown in (f). (f) Dispersion of the lollipop and peanut pockets, showing that the two pockets move through each other without any observable hybridization. This absence of an avoided crossing may be related to the different orbital character of the two pockets. Specifically, the lollipop pocket arises mostly from in-plane orbitals and the peanut pocket from out-of-plane orbitals. The lack of hybridization may be related to the $C_2$ symmetry of the (001) surface of NbP. (g) Equivalent of Fig. \ref{Fig3}d for TaAs. We see two co-propagating arcs with large spin-splitting due to the large SOC. Plot of the positions of the Weyl point projections in (h) NbP and (i) TaAs. We see that the separation of the Weyl points is $\sim 4$ times larger in TaAs due to the large SOC.}
\end{figure*}

\end{document}